\documentclass[12pt]{article}
\usepackage{amssymb,amsmath,cite,url}
\long\def\@makefntext#1{\parindent 0cm\noindent
\hbox to 1em{\hss$^{\@thefnmark}$}#1}

\begin{document}
\begin{titlepage}
\vspace{.5in}
\begin{flushright}
March 28, 2016\\
\end{flushright}
\vspace{.5in}
\begin{center}
{\Large\bf
Spontaneous  Dimensional Reduction\\[1ex]
in Quantum Gravity}\\
\vspace{.4in}
{S.~C{\sc arlip}\footnote{\it email: carlip@physics.ucdavis.edu}\\
       {\small\it Department of Physics}\\
       {\small\it University of California}\\
       {\small\it Davis, CA 95616}\\{\small\it USA}}
\end{center}

\vspace{.5in}
\begin{center}
{\large\bf Abstract}
\end{center}
\begin{center}
\begin{minipage}{4.7in}
{\small Hints from a number of different approaches to quantum gravity point to a
phenomenon of ``spontaneous dimensional reduction'' to two spacetime dimensions
near the Planck scale.  I examine the physical meaning of the term ``dimension'' in this
context, summarize the evidence for dimensional reduction, and discuss possible
physical explanations.
}
\end{minipage}
\vspace*{4ex}

{\small Essay written for the Gravity Research Foundation\\ 2016 Awards
 for Essays on Gravitation}
\end{center}
\end{titlepage}
\addtocounter{footnote}{-1}

The world of everyday experience has three dimensions of space
and one of time.  The idea that this could change at  small scales dates back a century 
\cite{Nordstrom,Kaluza,Klein}, and the notion  of ``extra dimensions'' is now commonplace 
in string theory and brane cosmology.  Recent research, though, hints at the opposite behavior, 
a \emph{reduction} in the number of dimensions near the Planck scale.  If true, this could require 
drastic revisions of our conception of spacetime, and might point toward new directions in 
quantum gravity.

\section*{Dimension as an observable}

To even pose the question of dimensional reduction, we must think carefully about the term
``dimension.''  In general relativity, spacetime is modeled as a smooth manifold,  and 
dimension  is unambiguous.  Kaluza-Klein theory uses higher-dimensional   
manifolds, again with no real ambiguity.  But quantum gravity is harder.  Already 
in quantum mechanics, path integrals are dominated by nowhere-smooth paths with 
fractal dimensions \cite{Morette}, and smooth one-dimensional paths appear only 
semiclassically.  If smooth large-scale spacetime is emergent, as many approaches to 
quantum gravity suggest, then dimension, too, must be emergent.

What, then, do we mean by dimension?  What we need are ``dimensional estimators,'' 
quantum observables with a simple dependence on dimension that can be generalized to situations 
in which the dimension is unclear.   Many possibilities exist, and  different choices need 
not always agree; dimension may depend on the precise question being asked.

Some dimensional estimators arise directly from geometry.  The simplest is the volume
 $V$ of a geodesic ball of radius $r$.  If, at some scale, $V\sim r^d$, then $d$ is a 
 measure of dimension at that scale.   The more general  Hausdorff dimension $d_H$  
counts the number of balls needed to cover a region as a function of $r$.  

Other estimators come from random walks or diffusion processes.  On a 
smooth manifold, the probability for a random walk from $x$ to reach $x'$ in time 
$s$ is given by a heat kernel,\footnote{$\sigma(x,x')$ is half the squared 
geodesic distance between $x$ and $x'$.}
\begin{equation}
K(x,x';s) \sim (4\pi s)^{-d/2} e^{-\sigma(x,x')/2s}
    \left( 1 + \mathcal{O}(s)\right)
\label{a1}
\end{equation}
The dependence on the dimension $d$ is universal, so $K(x,x';s)$ determines an 
effective dimension whenever a random walk can be defined.  In fact, there are
 two such dimensions: the ``walk dimension'' $d_W$, determined from mean   
distance as a function of time, and the spectral dimension $d_S$, determined from 
the ``return probability'' $K(x,x;s) \sim (4\pi s)^{-d_S/2}$.

These estimators assume a Riemannian (positive definite) metric, and their application 
to spacetimes requires analytic continuation, which might distort the 
physics \cite{Amelino}.  There are, however, several Lorentzian alternatives.  
The ``causal spectral dimension'' \cite{Eichhorn} exploits the relationship between two 
random walkers, each moving forward in time.  The Myrheim-Meyer 
dimension \cite{Myrheim,Meyer}  is built from volumes of causal diamonds.  We can 
also examine the behavior of geodesics \cite{Carlip1,Carlip2}: in certain spacetimes 
such as regions of Kasner space, although the manifold is four-dimensional, typical
geodesics  probe fewer dimensions \cite{Hu}.

We can augment these geometric probes with dimension-dependent physical phenomena.
This is an old idea: a century ago, Ehrenfest used such features as stability 
of orbits to single out three-dimensional space \cite{Ehrenfest}.  In thermodynamics,
free energy grows with temperature as $F/VT \sim T^{d-1}$, determining a ``thermodynamic 
dimension'' $d=d_T$; dimensional reduction in quantum gravity was first observed as a
scale-dependent change in this quantity \cite{Atick}.   Other thermodynamic dimensions 
can be built from such objects as the equation of state parameter and specific heat  
\cite{Amelino,Husain,Nozari}.

A final estimator comes from the fact that operators in quantum field theory acquire 
anomalous scaling dimensions, which change with energy under renormalization 
group flow.  In field theoretical approaches like asymptotic safety \cite{Weinberg}, 
these scaling dimensions determine a spacetime dimension.  In particular, two-point 
functions---propagators  and Greens functions---have a characteristic 
dimensional dependence: a massless Greens function $G(x,x')$ in $d$ dimensions 
goes as $\sigma(x,x')^{-(d-2)/2}$ (or $\ln \sigma(x,x')$ in two dimensions).  Quantum 
corrections then make $d$ a scale-dependent effective dimension.

\section*{Dimensional reduction}

With these estimators in hand, we can ask what quantum gravity says about   
spacetime dimension.  This is easier said than done: while we have promising 
research programs, we are still far from a complete quantum theory of gravity.   We may 
hope, though, that if a phenomenon like dimensional reduction occurs in enough different 
settings, it may point to a fundamental feature of the physics.

Scale dependence of dimension was originally noticed in anisotropic cosmologies 
\cite{Hu}, but its broader significance was not appreciated.  In  quantum 
gravity, the phenomenon first appeared in string theory, where a thermodynamic 
dimension was found to unexpectedly drop to $d_T=2$ at high temperatures,
leading Atick and Witten to postulate ``a lattice theory with a $(1+1)$-dimensional field 
theory on each lattice site''  \cite{Atick}. But it was only with the computation of spectral 
dimension in causal dynamical triangulations \cite{Ambjorn} that the idea really took 
hold.

Causal dynamical triangulations is a discrete, nonperturbative path integral approach to
quantum gravity in which spacetimes are approximated as piecewise flat simplicial 
complexes (higher dimensional ``geodesic domes'') with a prescribed direction of time.  
The model has been remarkably successful, yielding not only a de Sitter ground 
state but also a correct spectrum of quantum volume fluctuations \cite{Ambjorn2}.  
Lattice quantum gravity is notoriously bad at reproducing four-dimensional spacetime, 
though, and one must check the  dimension  carefully.  Using spectral dimension as 
their estimator, Ambj{\o}rn et al.\ found that causal dynamical triangulations 
has a phase in which spacetime is,  indeed, four-dimensional at large scales, but---%
much to their  surprise---appears two-dimensional at small scales.

At about the same time, Percacci and Perini pointed out that  if the renormalization group 
flow for quantum gravity has a  high energy (ultraviolet) fixed point---the hope of the asymptotic 
safety program---the scaling dimensions of fields at the fixed point are necessarily 
those of a two-dimensional field theory \cite{Percacci}.   The same dimensional reduction 
was confirmed for the spectral and walk dimensions \cite{Reuter}.  

These results sparked a broad investigation of small-scale dimension, with 
surprisingly consistent results.  Geometric and spectral dimension seem to fall to 
two at small scales for loop quantum gravity \cite{Modesto,Oriti} and spin foams 
\cite{Magliaro}.  For models with a minimum length, geometric \cite{Paddy} 
and spectral \cite{Modesto2} dimension do the same.  For noncommutative
Snyder space, thermodynamic dimensions show the same short-distance reduction 
\cite{Nozari}; for other noncommutative geometry models, the short-distance spectral 
dimension also decreases, although in a manner that depends on the choice of 
Laplacian \cite{Benedetti,Arzano}.  Models based on a generalized uncertainty principle 
show a reduction in thermodynamic dimension,  now to $d_T=2.5$ \cite{Husain}.  
The short-distance approximation of the Wheeler-DeWitt equation is dominated by 
spacetimes in which typical geodesics probe only two dimensions \cite{Carlip1,Carlip2}.   
In causal set theory, the naive spectral dimension increases at small scales \cite{Eichhorn}, 
but with a better choice of Laplacian it falls to  $d_S=2$ \cite{Belenchia}.  The 
Myrheim-Meyer dimension also falls to approximately two for small causal sets 
\cite{Carlip3}.  For proposed renormalizable modifications of general relativity, 
including Ho{\v r}ava-Lifshitz gravity \cite{Horava} and higher curvature models 
\cite{Calcagni}, a generalized spectral dimension again falls to $d_S=2$.
 
\section*{Extracting the physics}

By itself, no single one of these results is very convincing.  Together, though, they
show a clear pattern.  Like spontaneous symmetry breaking 
in field theory, dimensional reduction at short distances and high energies needs no
explicit external mechanism; it is indeed ``spontaneous dimensional reduction.''

But while this convergence suggests that the phenomenon is real, the underlying physical 
mechanism remains obscure.  The answer may require a full quantum theory of gravity,
but two interesting suggestions have been made:
\begin{enumerate}
\item{\bf Scale invariance:}  For general relativity to be ``asymptotic safe'' \cite{Weinberg}, 
its renormalization group flow must have an ultraviolet fixed point.  Such a fixed point is, 
by definition, scale-invariant, and this invariance is enough to guarantee effective 
two-dimensional behavior for any theory that includes general relativity \cite{Niedermaier}.  
While the details are technical, the underlying reason is simple: only in two dimensions is 
Newton's constant $G_N$ dimensionless and thus scale-free.

This is not completely satisfying, since it fails to explain \emph{why} such a fixed point 
should exist.  But perhaps the argument should be reversed: if we define the theory
at high energy, it is natural to postulate a high degree of symmetry, which is then broken
at lower energies.

\item{\bf Asymptotic silence:} Near a spacelike singularity, classical general relativity
exhibits ``asymptotic silence'': light cones shrink to lines, and nearby points become 
causally disconnected \cite{Uggla}.  This, in turn, leads to BKL behavior \cite{BKL}, in which
the metric is locally Kasner with axes of anisotropy that vary chaotically in space and time.  
In such a spacetime, each point has a ``preferred'' spatial direction, and geodesics effectively 
see only $1+1$ dimensions \cite{Carlip1,Carlip2}.  

While this behavior was discovered in cosmology, asymptotically silent spacetimes dominate
the short-distance Wheeler-DeWitt equation, and  quantum fluctuations near the Planck scale 
may lead to short-distance asymptotic silence everywhere \cite{Pitelli}.  If so, the small scale
 metric will have two length scales,
\begin{equation}
ds^2 = \ell_\parallel^2g_{\mu\nu}dx^\mu dx^\nu + \ell_\perp^2 h_{ij}dx^idx^j
\label{a2}
\end{equation}
and the Einstein-Hilbert action will become very nearly that of a two-dimensional conformal
field theory for the transverse metric $h_{ij}$ \cite{Carlip4}.  Such reductions have been studied 
in different contexts \cite{tHooft,Verlinde,Kabat}, and could give a physical explanation for
the fixed point in asymptotic safety.
\end{enumerate}
It is too soon to tell whether either, or perhaps both, of these ideas will explain spontaneous 
dimensional reduction.  But it seems fair to say that the phenomenon offers  an exciting 
new glimpse into the decades-old problem of quantize gravity.

\vspace{1.5ex}
\begin{flushleft}
\large\bf Acknowledgments
\end{flushleft}

This work was supported in part by Department of Energy grant
DE-FG02-91ER40674.


\begin{thebibliography}{99}
\bibitem{Nordstrom} G.\ Nordstr{\"o}m, Phys.\ Z.\ 15 (1914) 504; English translation at
    arXiv:physics/0702221. 
\bibitem{Kaluza} T.\ Kaluza, Sitzungsber.\ Preuss.\ Akad.\ Wiss.\ Berlin (Math.\ Phys.) K1
    (1921) 966.
\bibitem{Klein} O.\ Klein, Z.\ Phys.\  37 (1926) 895.
\bibitem{Morette} P.\ Cartier and C.\ DeWitt-Morette, \emph{Functional Integration}, Cambridge
      University Press, 2006.
\bibitem{Amelino} G.\ Amelino-Camelia, F.\ Brighenti, G.\ Gubitosi, and G.\ Santos,
      arXiv:1602.08020.
\bibitem{Eichhorn} A.\ Eichhorn and S.\ Mizera, Class.\ Quant.\ Grav. 31 (2014) 125007,
      arXiv:1311.2530.
\bibitem{Myrheim} J.~Myrheim, 1978 CERN preprint TH-2538.
\bibitem{Meyer} D.~A.~Meyer, Ph.D.\ thesis, MIT (1989), \url{http://hdl.handle.net/1721.1/14328}.
\bibitem{Carlip1} S.\ Carlip, in \emph{Proc.\ of the 25th Max Born Symposium: The Planck 
     Scale}, AIP Conf.\ Proc.\ 1196 (2009) 72,  arXiv:1009.1136.
\bibitem{Carlip2} S.\ Carlip, in \emph{Foundations of Space and Time}, edited by J.\ Murugan, 
      A.\ Weltman, and G.~F.~R.\ Ellis, Cambridge University Press, 2012, arXiv:1009.1136.
\bibitem{Hu} B.~L.\ Hu and D.~J.\ O'Connor, Phys.\ Rev.\ D34 (1986) 2535. 
\bibitem{Ehrenfest} P.\ Ehrenfest, KNAW Proc.\ 20 I (1918) 200.
\bibitem{Atick} J.~J.\ Atick and E.\ Witten,  Nucl.\ Phys. B310 (1988) 291.
\bibitem{Husain} V.\ Husain, S.~S.\ Seahra, and E.~ J.\ Webster, Phys.\ Rev.\ D88 (2013)
024014, arXiv:1305.2814.
\bibitem{Nozari}  K.\ Nozari, V.\ Hosseinzadeh, and M.~A.\ Gorji, Phys.\ Lett.\ B750 
       (2015) 218, arXiv:1504.07117.
\bibitem{Weinberg} S.\ Weinberg, in \emph{General Relativity: An Einstein Centenary Survey}, 
       edited by S.~W.\ Hawking and W.\ Israel, Cambridge University Press, Cambridge, 1979.
\bibitem{Ambjorn} J.~Ambj{\o}rn, J.~Jurkiewicz, and R.~Loll,  Phys.\ Rev.\ Lett.\
   95 (2005) 171301, arXiv:hep-th/0505113.
\bibitem{Ambjorn2} J.\ Ambj{\o}rn, A.\ G{\"o}rlich, J.\ Jurkiewicz, R.\ Loll, J.\ Gizbert-Studnicki, and
       T.\ Trz{\'e}sniewski, Nucl.\ Phys.\ B849 (2011) 144, arXiv:1102.3929.
\bibitem{Percacci} R.~Percacci and D.~Perini, Class.\ Quant.\ Grav.\ 21 (2004) 
   5035, arXiv:hep-th/0401071.
\bibitem{Reuter} M.~Reuter and F.~Saueressig, JHEP 1112 (2011) 012,  arXiv:1110.5224.
\bibitem{Modesto} L.~Modesto, Class.\ Quant.\ Grav.\ 26  (2009) 242002, arXiv:0812.2214.
\bibitem{Oriti} G.\ Calcagni, D.\ Oriti, and J.\ Th{\"u}rigen, Phys.\ Rev.\ D 91, 084047 (2015).
    arXiv:1412.8390.
\bibitem{Magliaro} E.\ Magliaro, C.\ Perini, and L.\ Modesto, arXiv:0911.0437.
\bibitem{Paddy} T.\ Padmanabhan, S.\ Chakraborty, and D.\ Kothawala, arXiv:1507.05669.
\bibitem{Modesto2} L.\ Modesto and P.\ Nicolini, Phys.\ Rev.\ D81 (2010) 104040,
     arXiv:0912.0220.
\bibitem{Benedetti} D.~Benedetti, Phys.\ Rev.\ Lett.\ 102 (2009) 111303, arXiv:0811.1396.
\bibitem{Arzano} M.~Arzano and T.~Trzesniewski, Phys.\ Rev.\ D89 (2014) 124024, 
     arXiv:1404.4762.
\bibitem{Belenchia} A.\ Belenchia, D.~M.~T.\ Benincasa, A.~Marciano,and L.~Modesto, 
     Phys.\ Rev.\ D93 (2016) 044017, arXiv:1507.00330.
\bibitem{Carlip3}  S.\ Carlip, Class.\ Quant.\ Grav.\ 32 (2015) 232001, arXiv:1506.08775.
\bibitem{Horava}  P.\ Ho{\v r}ava, Phys.\ Rev.\ Lett.\ 102 (2009) 161301, arXiv:0902.3657.
\bibitem{Calcagni} G.\ Calcagni, L.\  Modesto, and G.\  Nardelli, arXiv:1408.0199.
\bibitem{Niedermaier} M.\ Niedermaier and M.\ Reuter, Living Rev. Relativity 9 (2006) 5,
       URL (cited March 2016): \url{http://www.livingreviews.org/lrr-2006-5}.
\bibitem{Uggla} J.~M.\ Heinzle, C.\ Uggla, and N.\ R{\"o}hr, Adv.\ Theor.\ Math.\ Phys.\ 13 (2009)
       293, arXiv:gr-qc/0702141.
\bibitem{BKL} V.~A.\ Belinskii, I.~M.\ Khalatnikov, and E.~M.\ Lifshitz, Adv.\ Phys. 19  (1970) 525; 
       Adv.\ Phys.\ 31 (1982) 639.
\bibitem{Pitelli} S.\ Carlip, R.~A.\ Mosna, and J.~P.~M.\ Pitelli, Phys.\ Rev.\ Lett.\ 107, (2011)
     021303, arXiv:1103.5993.
\bibitem{Carlip4} S.\ Carlip, AIP Conf.\ Proc.\ 1483 (2012) 63, arXiv:1207.4503.
\bibitem{tHooft} G.\  't Hooft, Phys.\ Lett.\ 198B (1987) 61.
\bibitem{Verlinde} H.~L.\ Verlinde and E.~P.\ Verlinde, Nucl.\ Phys.\ B371(1992) 246, 
     arXiv:hep-th/9110017.
\bibitem{Kabat} D.\ Kabat and M.\ Ortiz, Nucl.\ Phys.\ B388 (1992) 570, arXiv:hep-th/9203082.
 
\end{thebibliography}
\end{document}